\documentclass[twoside]{ilcws07}
\usepackage[latin1]{inputenc}
\usepackage[dvips]{graphicx,epsfig,color}
\usepackage{wrapfig,rotating}
\usepackage{amssymb,amsmath,array}

\begin{document}
\title{
Light Pseudoscalars in Little Higgs Models at the ILC} 
\author{J\"urgen Reuter
\vspace{.3cm}\\
Albert-Ludwigs-Universit\"at Freiburg -  Physikalisches Institut \\
Hermann-Herder-Str.~3, D-79104 Freiburg - Germany
}

\maketitle

\begin{abstract}
  We discuss the properties of light pseudoscalar particles, the
  so-called pseudoaxions, within Little Higgs models, focusing on
  their phenomenology at the ILC. We especially discuss a method of
  how to distinguish between the two basic classes of Little Higgs
  models, the product and simple group models, by a specific
  production channel and decay mode. These are strictly forbidden in
  the product group models.
\end{abstract} 
 
\section{Little Higgs Models}

Little Higgs Models~\cite{lhm_moose} provide a
solution to the hierarchy problem, as they stabilize the Higgs boson
against quadratic divergences at the one-loop level by the mechanism
of collective symmetry breaking: the Higgs is charged under two global 
symmetry groups, which both need to be broken in order to lift the
flat direction in the potential of the Higgs boson and make it a
pseudo-Nambu-Goldstone boson (PNGB). During the last years a
bewilderment of different models has been developed. 

\begin{wrapfigure}{l}{0.5\columnwidth} 
\centerline{\includegraphics[width=0.45\columnwidth,height=.39\columnwidth]{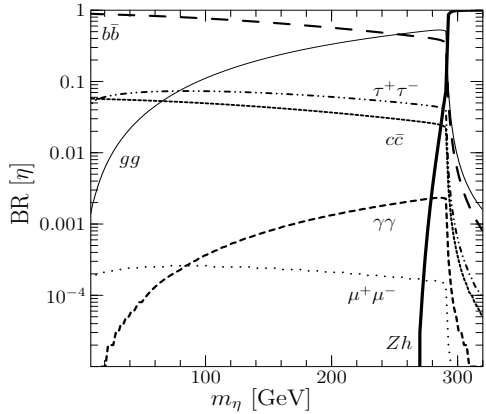}}  
\caption{Branching ratios of the pseudoaxion in the Simplest Little
  Higgs as a function of its mass.}\label{fig:etabr}  
\end{wrapfigure} 
These models can be classified in three different categories, the
so-called moose 
models with a moose diagram structure of links of global and local
symmetry groups, the product-group models and the simple-group
models. In the product-group models (the most-studied case is the
Littlest Higgs) the electroweak gauge group is doubled, broken down to
the group $SU(2)_L$, while the Higgs shares together with the other
PNGBs an irreducible representation of the coset space of the global
symmetry breaking. On the other hand, in simple-group models the
electroweak gauge group is enlarged to a simple $SU(N)$ group, while
the Higgs is distributed over several multiplets of the global
symmetry group, which usually has a product group structure similar to
chiral symmetries in QCD~\cite{simple}. For an overview,
see~\cite{review}. 

The two crucial scales in the Little Higgs set-up are the cut-off
scale $\Lambda$ where the models are embedded in a UV-complete theory
(usually a strongly-interacting theory with a partonic substructure of
the PNGBs) and the intermediate scale $F$ which determines the masses
and decay constants of the PNGBs (except for the Higgs which is down
at $v$ by the collective symmetry breaking mechanism). Electroweak
precision observables and direct search limits~\cite{lowenergy} tell
us that the scale $F$ must be at least of the order of $1-2$
TeV. Paradoxically, the Higgs boson in Little Higgs models tends to be
quite heavy compared to the Standard Model or the MSSM, of the order
of $200-600$ GeV~\cite{little_kr}. For Little Higgs model scales that
high most new particles will be produced close to the kinematical limit at
the LHC, such that a precision determination of their parameters might
be difficult. Furthermore, also the sensitivity of the ILC in indirect
measurements might be limited, if the new phyics does couple to SM
fermions only very weakly~\cite{resonances}. A method to distinguish
between different models, especially at the LHC, is highly
welcome. Such a method will be presented here.   

\begin{wrapfigure}{r}{0.5\columnwidth} 
\centerline{\includegraphics[width=0.45\columnwidth,height=.39\columnwidth]{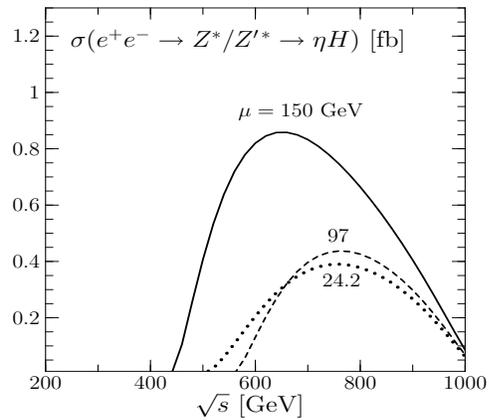}} 
\caption{Cross section for the $H\eta$ associated production at ILC,
  taking into account the destructive $Z/Z'$ interference. The full,
  dashed and dotted lines correspond to $m_\eta = 309/200/50$ GeV,
  respectively.}\label{fig:xsec}  
\end{wrapfigure} 


\section{Pseudoaxions in Little Higgs models}

Little Higgs models generally have a huge global symmetry group, which
contains not only products of simple groups but also a certain number
of $U(1)$ factors. These Abelian groups can either be gauged, in which
case they lead to a $Z'$ boson, or they are only (approximate) global
symmetries. In the latter case there is a pseudo-Goldstone boson
attached to that spontaneously broken global $U(1)$
factor~\cite{pseudoaxions}. The number of pseudoaxions in a given
model is determined by the mismatch between the rank reduction in
the global and the local symmetry group, since it gives the number of
uneaten bosons. In the Littlest Higgs, e.g., there is one such
pseudoaxion, in the Simplest Little Higgs~\cite{simplest} there is
one, in the original simple group model there are two, in the minimal
moose model there are four, and so on.

These particles are electroweak singlets, hence all couplings to SM
particles are suppressed by the ratio of the electroweak over the
Little Higgs scale, $v/F$. There mass lies in the range from several
GeV to a few hundred GeV, being limited by a naturalness argument and
the stability of the Coleman-Weinberg potential. For the Simplest
Little Higgs, on whose phenomenology we will concentrate here, there
is a seesaw between the Higgs and the pseudoscalar
mass~\cite{pseudoaxions}, determined by the explicit symmetry breaking
parameter $\mu$, where $m_\eta \approx \sqrt{2} \mu$.  Since the
pseudoaxions inherit the Yukawa coupling 
\begin{figure} 
\centerline{\includegraphics[width=0.45\columnwidth]{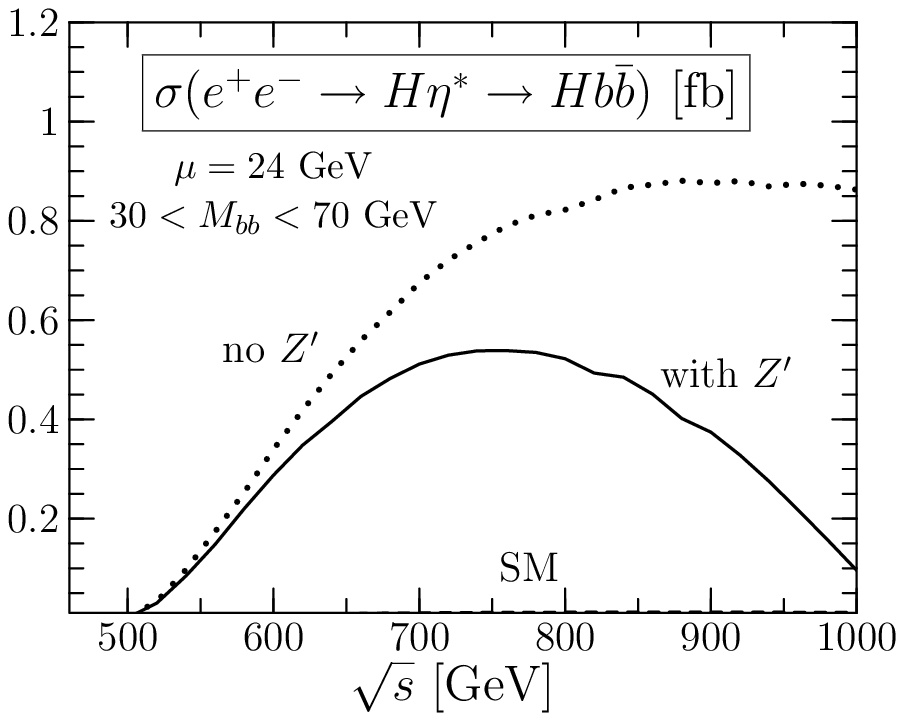}
\includegraphics[width=0.45\columnwidth]{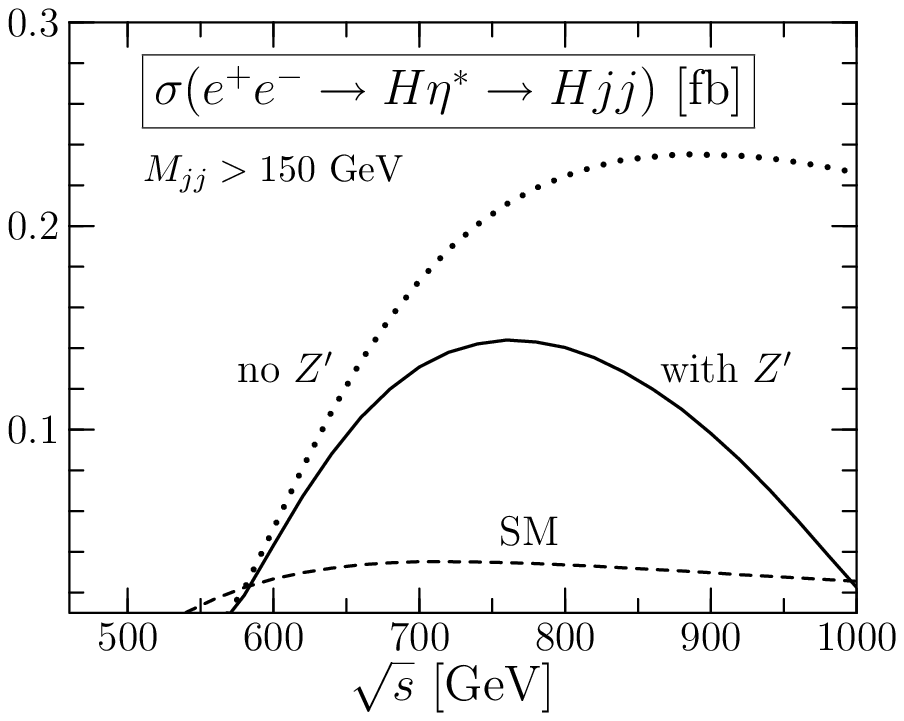}
} 
\caption{ILC cross section, left: small-mass $\eta$ with final state
  $Hbb$, right intermediate mass with final state $Hgg$. The dotted
  line is without the $Z/Z'$ interference, the dashed one the SM
  background.} 
  \label{fig:hbbhjj}  
\end{figure} 
structure from the Higgs bosons, they decay predominantly to the
heaviest available fermions in the SM, and because of the absence of
the $WW$ and $ZZ$ modes, the anomaly-induced decays $gg$ and
$\gamma\gamma$ are sizable over a wide mass range,
cf.~Fig.~\ref{fig:etabr}. From this, one can see that as soon as the
decay to $HZ$ is kinematically allowed, it dominates completely. Such
a $\eta HZ$ coupling, which is possible only after electroweak
symmetry breaking and hence proportional to $v/F$, is only allowed in
simple group models and is forbidden to all orders in product group
models. One can factor out the $U(1)_\eta$ group from the matrix of
pseudo-Goldstone bosons. We use $\xi=\exp\left[i\eta/F\right]$ for the
pseudo-axion field and $\Sigma=\exp\left[i\Pi/F\right]$ for the
non-linear representation of the remaining Goldstone multiplet $\Pi$
of Higgs and other heavy scalars.  Then, for product group models, the
kinetic term may be expanded as
\begin{align}
  \label{LL-kin}
  \mathcal{L}_{\text{kin.}} \sim F^2
  \text{Tr} \left[ (D^\mu (\xi \Sigma)^\dagger (D_\mu (\xi \Sigma)) \right]
  &= \ldots - 2F(\partial_\mu \eta)\,
  \text{Im} \text{Tr} \left[  (D^\mu \Sigma)^\dagger \Sigma \right]
  + O(\eta^2),
\end{align}
where we write only the term with one derivative acting on $\xi$ and
one derivative acting on $\Sigma$.  This term, if nonzero, is the only
one that can yield a $ZH\eta$ coupling.

We now use the special structure of the covariant derivatives in
product group models, which is the key to the Little Higgs mechanism:
$D_\mu \Sigma = \partial_\mu \Sigma + A_{1,\mu}^a \left( T_1^a \Sigma 
+ \Sigma (T_1^a)^T \right) + A_{2,\mu}^a \left( T_2^a \Sigma
+ \Sigma (T_2^a)^T \right)$, 
where $T_i^a, i=1,2$ are the generators of the two independent $SU(2)$
groups, and $A_{i,\mu}^a = W^a_\mu$ + heavy fields. Neglecting the heavy
gauge fields and extracting the electroweak gauge bosons, we have
$\text{Tr}\left[ (D^\mu \Sigma)^\dagger \Sigma \right] \sim W_\mu^a
\text{Tr} \left[ (T_1^a + T_2^a) + (T_1^a + T_2^a)^* \right] = 0$. 
This vanishes due to the zero trace of $SU(2)$ generators.  The same
is true when we include additional $U(1)$ gauge group generators such
as hypercharge, since their embedding in the global simple group
forces them to be traceless as well.  We conclude that the coefficient
of the $ZH\eta$ coupling vanishes to all orders in the $1/F$
expansion.

Next, we consider the simple group models, where we
use the following notation for the nonlinear sigma fields:
$\Phi\zeta$, where $\Phi=\exp[i\Sigma/F]$ and $\zeta=\left(0,\ldots
0,F\right)^T$ is the vev directing in the $N$ direction for an $SU(N)$
simple gauge group extension of the weak group.  Thus, in simple group
models the result is the $N,N$ component of a matrix:
\begin{equation}\label{kin:simplegroup}
\mathcal{L}_{\text{kin.}} \sim F^2 D^\mu (\zeta^\dagger \Phi^\dagger) D_\mu
 (\Phi\zeta) = \ldots +  i F (\partial_\mu \eta) \left(
 \Phi^\dagger (D_\mu \Phi) - (D_\mu \Phi^\dagger) \Phi \right)_{N,N} \; .
\end{equation}
We separate the last row and column in the matrix representations of
the Goldstone fields $\Sigma$ and gauge boson fields $\mathbf{V}_\mu$:
the Higgs boson in simple group models sits in the off-diagonal
entries of $\Sigma$, while the electroweak gauge bosons reside in the
upper left corner of $\mathbf{V}_\mu$. With the
Baker-Campbell-Hausdorff identity, one gets for the term in
parentheses in Eq.~(\ref{kin:simplegroup}): 
\begin{multline}\label{eq:vexpansion}
  \mathbf{V}_\mu + \frac{i}{F} \lbrack \Sigma, \mathbf{V}_\mu \rbrack 
  - \frac{1}{2F^2} \lbrack \Sigma , \lbrack \Sigma, \mathbf{V}_\mu
  \rbrack \rbrack + \ldots \\
  =\begin{pmatrix}
      \mathbf{W}_\mu & 0 \\ 
      0 & 0 
   \end{pmatrix}
   + \frac{i}{F} 
  \begin{pmatrix}
   0 & - \mathbf{W}_\mu h \\ h^\dagger \mathbf{W}_\mu & 0 
  \end{pmatrix} - \frac{1}{2F^2} 
  \begin{pmatrix}
    h h^\dagger \mathbf{W} + \mathbf{W} h h^\dagger & 0
    \\ 0 & - 2 h^\dagger \mathbf{W} h 
  \end{pmatrix} + \ldots
\end{multline}
The $N,N$ entry can only be nonzero from the third term on. The first
term, would be a mixing between the $\eta$ and the Goldstone boson(s)
for the $Z'$ state(s) and cancels with the help of the many-multiplet
structure. If the $N,N$ component of the second term were nonzero, it
would induce a $ZH\eta$ coupling without insertion of a factor $v$.
This is forbidden by electroweak symmetry.  To see this, it is
important to note that in simple group models the embedding of the
Standard Model gauge group always works in such a way that hypercharge
is a linear combination of the $T_{N^2-1}$ and $U(1)$ generators.
This has the effect of canceling the $\gamma$ and $Z$ from the
diagonal elements beyond the first two positions, and preventing the
diagonal part of $\mathbf{W}_\mu$ from being proportional to $\tau^3$.
The third term in the expansion yields a contribution to the $ZH\eta$
coupling, $(\partial^\mu\eta)h^\dagger \mathbf{W}_\mu h \sim v H 
Z_\mu\partial^\mu\eta$.

\begin{wrapfigure}{r}{.5\columnwidth}
\centerline{\includegraphics[width=0.45\columnwidth]{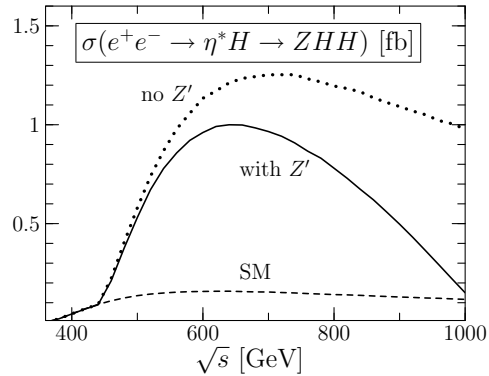}} 
\caption{ILC cross section for a high-mass $\eta$ with final state
  $ZHH$. The dotted line is without the $Z/Z'$ interference, the
  dashed one the SM background.}\label{fig:zhh}   
\end{wrapfigure} 

The crucial observation is that the
matrix-representation embedding of the two non-Abelian $SU(2)$ gauge
groups, and especially of the two $U(1)$ factors within the
irreducible multiplet of the pseudo-Goldstone bosons of one simple
group (e.g. $SU(5)$ in the Littlest Higgs), is responsible for the
non-existence of this coupling in product group models.  It is exactly
the mechanism which cancels the quadratic one-loop divergences
between the electroweak and heavy $SU(2)$ gauge bosons which cancels
this coupling.  In simple group models the Higgs mass term
cancellation is taken over by enlarging $SU(2)$ to $SU(N)$, and the
enlarged non-Abelian rank structure cancels the quadratic divergences
in the gauge sector -- but no longer forbids the $ZH\eta$
coupling. Hence, its serves as a discriminator
between the classes of models.  


\section{ILC phenomenology}

The pseudoaxion can be discovered at the LHC in gluon fusion and
observed in the rare decay mode
$\gamma\gamma$~\cite{pseudoaxions}. But the $\eta HZ$ coupling can be
observed at the LHC only if either one of the decays $H \to 
Z \eta$ or $\eta \to ZH$ is kinematically allowed. This leaves large
holes in parameter space, which can be covered by a $500-1000$ GeV
ILC, depending on the masses. Here, we focus on the discovery
potential of the ILC for the pseudoaxions, assuming the presence of
the $ZH\eta$ coupling. We focus on the Simplest Little Higgs with
parameters chosen to fulfill the low-energy constraints. The
production happens via an $s$-channel $Z$ exchange, in association with a
Higgs boson like in a two-Higgs-model. Fig.~\ref{fig:xsec} shows the
cross section as a function of $\sqrt{s}$ for three
different values of the $\eta$ mass. The simulations for the processes
discussed here have been performed with the {\sc whizard/omega}
package~\cite{omega,whizard,omwhiz}, which is ideally suited for
physics beyond the SM~\cite{omwhiz_bsm}. In the following, we assume that
the Higgs properties are already known from LHC, and that the Higgs
can be reconstructed. In Fig.~\ref{fig:hbbhjj} and~\ref{fig:zhh} we
show the three different possible final states, depending on the
dominant branching ratio of the pseudoaxion: for low masses (up to
approx.~150 GeV) this is $b\bar b$, in an intermediate range (between
150 GeV and 270 GeV) the $gg$ (hence dijet) mode is largest, while
$ZH$ sets in above masses of 270 GeV. The figures show the effect of a
destructive $Z/Z'$ interference, which brings cross sections down by a
factor of two at the peak, but never endangers visibility. SM
backgrounds are nowhere an issue, from marginal for $Hjj$ to
negligible in the $Hbb$ case. Interesting is the $ZHH$ final state
which is important for measuring the triple Higgs
coupling~\cite{triplehiggs}. In the SM the cross section is at the
borderline of detectability, but Fig.~\ref{fig:zhh} shows that rates
are larger by factors two to six in the Simplest Little Higgs with the
intermediate pseudoaxion. In conclusion, the ILC provides an ideal
environment for discovering pseudoaxions and measuring their
properties. The $ZH\eta$ coupling provides tool for the discrimination
between simple and product group models.
 
\section{Acknowledgments} 
 
JR was partially supported by the Helmholtz-Gemeinschaft under
Grant No. VH-NG-005.

 
\section{Bibliography} 
 
\begin{footnotesize} 

\end{footnotesize}
 
 
\end{document}